\documentclass{article}
\usepackage[utf8]{inputenc}
\usepackage{graphicx}
\usepackage{subcaption}
\usepackage{caption}
\usepackage{amsmath}
\usepackage{hyperref}
\usepackage{xargs}      
\usepackage{booktabs}
\usepackage{url}
\usepackage[pdftex,dvipsnames]{xcolor}  % Coloured text etc.
\usepackage[colorinlistoftodos,prependcaption,textsize=tiny]{todonotes}
\newcommandx{\unsure}[2][1=]{\todo[linecolor=red,backgroundcolor=red!25,bordercolor=red,#1]{#2}}
\newcommandx{\change}[2][1=]{\todo[linecolor=blue,backgroundcolor=blue!25,bordercolor=blue,#1]{#2}}
\newcommandx{\info}[2][1=]{\todo[linecolor=OliveGreen,backgroundcolor=OliveGreen!25,bordercolor=OliveGreen,#1]{#2}}

\title{\textbf{Verifying a Cruise Control System using Simulink and SpaceEx}}
\author{\textbf{Nikolaos Kekatos}\\~\\ \emph{Verimag Laboratory, University of Grenoble Alpes}\\
 \href{mailto:nikolaos.kekatos@univ-grenoble-alpes.fr}{nikolaos.kekatos@univ-grenoble-alpes.fr}
 }
\date{}

\begin{document}

\maketitle

\begin{abstract}
    This article aims to provide a simple step-by-step guide highlighting the steps needed to verify a control system with formal verification tools. Starting from a description of the physical system and a control objective in natural language, we design the plant and the controller, we use Simulink for simulation and we employ a reachability analysis tool, SpaceEx, for formal verification. 
\end{abstract}
\section{Introduction}

The goal of this article is to model a simple control system~\cite{lee2006cyber}, conduct reachability analysis and verify given control requirements. In this work, we consider a \textit{Cruise Control} system of an automotive vehicle. Cruise control is installed on most modern vehicles and its purpose is to regulate and maintain the vehicle speed despite external disturbances, such as changes in wind or road grade. This is accomplished by measuring the vehicle speed, comparing it to the desired or reference speed, and automatically adjusting the throttle according to a control law~\cite{kesting2008adaptive}.

The rest of this paper is organized as follows. Section \ref{model} presents the physical model (plant) of the vehicle. Section \ref{obj} establishes the control requirements and Section \ref{Control} introduces the control design part. Section \ref{Simulink} presents simulation results the Simulink model and Section \ref{SpaceEx} shows  reachability results with SpaceEx.

\section{Modeling}\label{model}

\begin{figure}[h!]
	\centering
\includegraphics[width=0.6\textwidth]{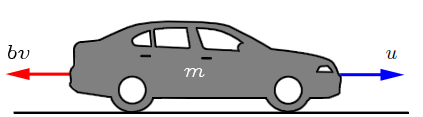}
	\caption{Free body diagram}
	\label{fig:car}
\end{figure}

In this section, a simplified model of the vehicle dynamics is presented. A free-body diagram of the dynamics is shown in Figure \ref{fig:car}.

The vehicle's mass  is described by $m$. The control force $u$ represents the force generated at the road/tire interface. For this simplified model, it is assumed that the force can be controlled directly, while the powertrain and tire dynamics are neglected. The resistive forces, $bv$, due to rolling resistance and wind drag, are assumed to vary linearly with the vehicle velocity, $v$, and act in the direction opposite the vehicle's motion.

Based on Newton law, the system equation is as follows.
\begin{align}
m\dot{v}+bv&=u
\end{align}

It is considered that both the vehicle speed and acceleration can be measured. The state space representation is presented below.% \change{Not really necessary. We could ignore this part.}
\begin{align*}
\dot{x}&=v\\
\dot{v}&=-\frac{b}{m}v+\frac{1}{m}u
\end{align*}
The equivalent state-space representation follows.%, \change{This is a matrix notation. It could be ignored too.}

\[
\left[
\begin{array}{c}
\dot{x}\\
\dot{v}
\end{array}
\right]
=
\left[
\begin{array}{c c}
0& 1\\
0 & -\frac{b}{m} 
\end{array}
\right]
\left[
\begin{array}{c}
x\\
v  
\end{array}
\right]
+
\left[
\begin{array}{c}
0\\
\frac{1}{m}  
\end{array}
\right]
u 
\]

\section{Requirements} \label{obj}

As mentioned in the previous section, the fundamental objective of a cruise controller is to regulate the vehicle's speed in accordance with the road conditions and road participants. In this work, we consider that the vehicle is part of a vehicle platoon and there is a leading vehicle ahead of it. As such, the desired vehicle speed is communicated by the leading vehicle and it can be seen as an external input to our system.  

Cruise control systems entail several control objectives/requirements. In the context of this work, we consider two of the most critical ones. The first is to regulate the vehicle speed so that the vehicle can eventually reach the reference/desired speed. To make this requirement more realistic, we enlarge the desired speed range by a percentage. This percentage refers to the "steady-state error" and is typically between 2\% to 5\%. The second requirement is related to the time that the vehicle needs to reach the desired speed range. We use the control theoretic notation of "rise-time". Rise-time corresponds to the time required by a signal/variable (vehicle speed for our case) to reach the 90\% of the desired speed. %For instance, if the desired speed is $20$m/s, the rise-time relates to the time that the vehicle speed reaches 18m/s. 
We could also consider requirements related to the distance, acceleration, passenger comfort.
\newpage

\section{Control Design} \label{Control}

The parameters of
the vehicle and control requirements are given in Table \ref{table:param}. They are taken from \cite{ref}.

\begin{table}[ht]
\centering
\begin{tabular}{@{}c c c@{}}
\toprule {Parameters} & {Values} & {Units} \\ \midrule% centered columns (4 columns) 
 % inserts single horizontal line 
$b$ & 50 & $N s/m$\\ % inserting body of the table 
$m$ & 1000 & $kg$\\ 
initial speed & 0 & $m/s$\\
desired speed & 15 & $m/s$\\
rise time & $<5$ & $s$\\
steady-state error & $<5$\% & -\\ \midrule
\end{tabular} 
\caption{Vehicle parameters}
\label{table:param} % is used to refer this table in the text 
\end{table}

\begin{figure}[ht]
    \centering
    \includegraphics[width=1\textwidth]{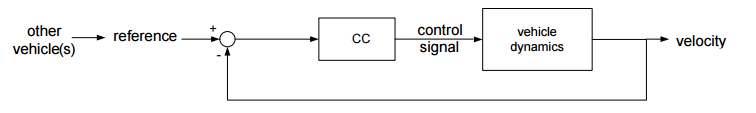}
    \caption{Block diagram of a vehicle equipped with cruise control}
    \label{fig:imag}
\end{figure}

Figure \ref{fig:imag} presents a high level picture of a Cruise Control system, pinpointing the input and the output.

%\subsection{Vehicle Parameters}

%\subsection{Abstraction}

%The physical variables are bounded in practice. For this example, the following assumptions are considered.

% \begin{table}[h!]
% \centering % used for centering table 
% \begin{tabular}{c c c} % centered columns (4 columns) 
% \hline\hline %inserts double horizontal lines 
% Variable & Minimum & Maximum \\ [0.5ex] % inserts table 
% %heading 
% \hline % inserts single horizontal line 
% $x$ & 0 & 400\\ % inserting body of the table 
% $v$ & -5 & 20\\ 
% $a$ & -6 & 6 \\
% $u$ & -4000 & 4000\\
%  [1ex] % [1ex] adds vertical space 
% \hline %inserts single line 
% \end{tabular}
% \caption{ Range of physical variables}
% \label{table:nonlin} % is used to refer this table in the text 
% \end{table}

The option of control method aligns with the common practices in automotive sector and its applicability with SpaceEx. In this respect, a PID controller is considered. In particular, we opt for a PI controller with gains $K_P=800$ and $K_I=40$. %A simulation result follows.

\section{Simulations with Simulink} \label{Simulink}

Simulink is a graphical programming environment for modeling, simulating and analyzing dynamical systems. It includes a set of block libraries and is a commonly used tool in industry.

\begin{figure}[h!]
	\centering
\hspace*{-1em}\includegraphics[scale=0.7]{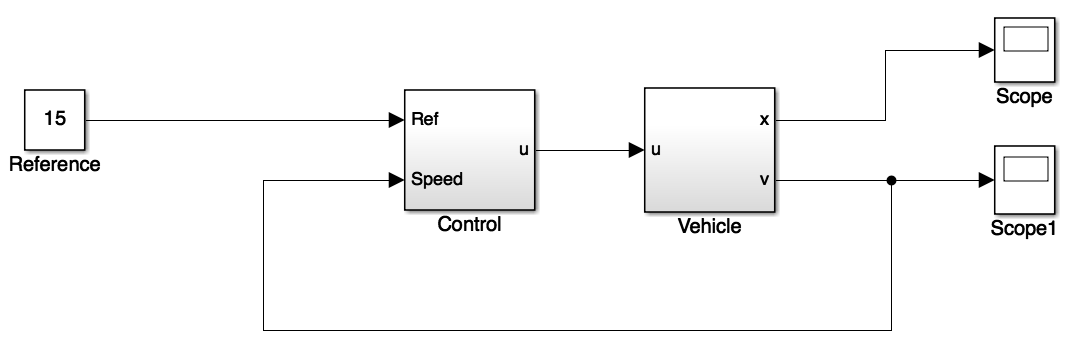}
	\caption{Cruise Control - Top Level - Simulink}
	\label{fig:car1}
\end{figure}

\begin{figure}[h!]
	\centering
\hspace*{-2em}\includegraphics[scale=0.8]{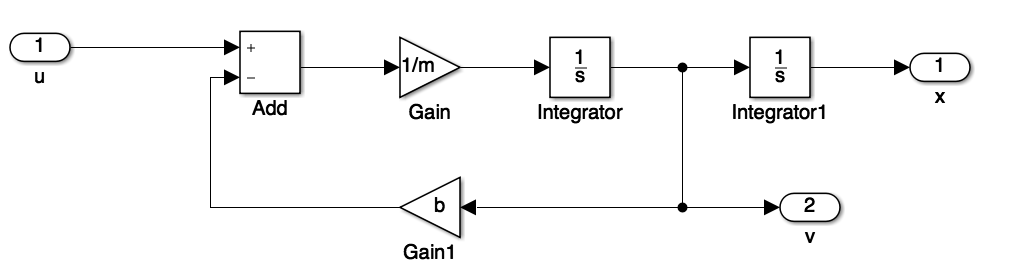}
	\caption{Cruise Control - Vehicle Dynamics - Simulink}
	\label{fig:car2}
\end{figure}

\begin{figure}[h!]
	\centering
\includegraphics[scale=0.7]{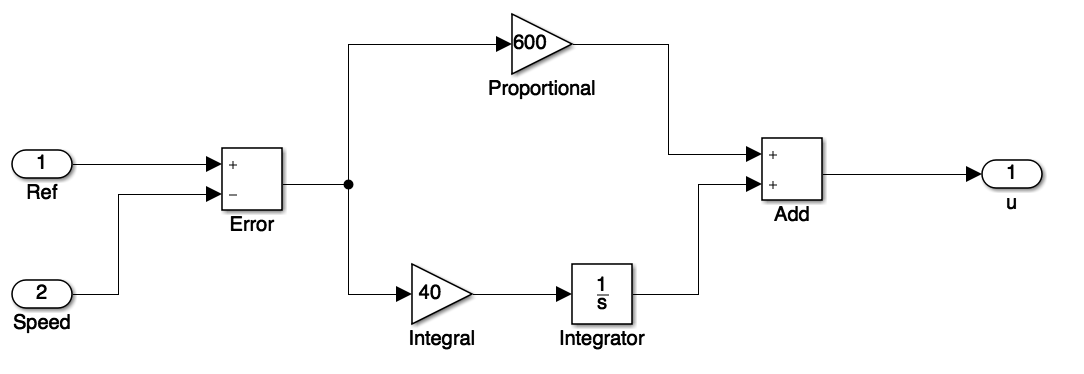}
	\caption{Cruise Control - Controller - Simulink}
	\label{fig:car3}
\end{figure}

The Simulink model (\texttt{CC\_Simulink.slx}) of the cruise controller along with its subsystems  are portrayed in Figures \ref{fig:car1}, \ref{fig:car2} and \ref{fig:car3}. The model files can be found at Github\footnote{The url is \url{https://github.com/nikos-kekatos/SpaceEx-tutorials}.}.
\newpage

\begin{figure}[ht]
	\centering
\includegraphics[scale=0.3]{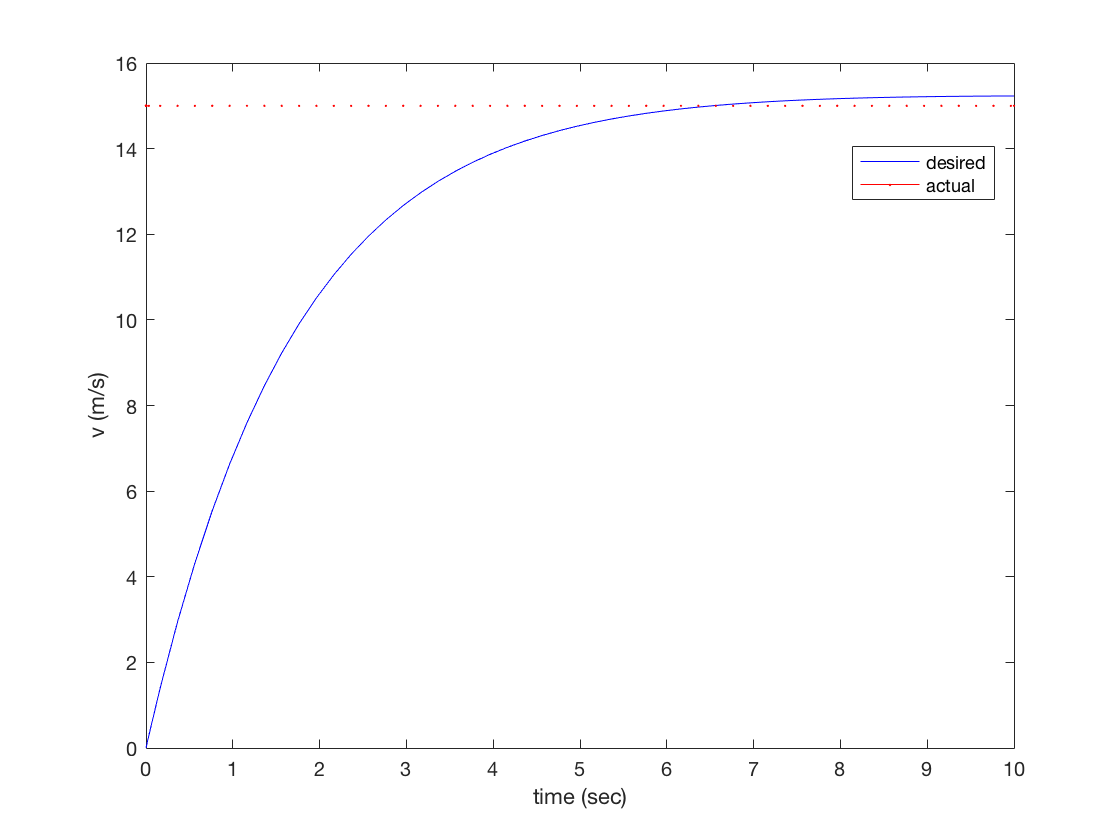}
	\caption{Cruise Control - Simulation with Simulink - Initial speed: 0 m/s, desired speed: 15 m/s}
	\label{fig:simulation}
\end{figure}

\newpage
Considering that the vehicle starts from idle and the objective is to reach 15~m/s, we get the simulation results  shown in figure \ref{fig:simulation}.

\section{Verification with SpaceEx}
\label{SpaceEx}

\begin{figure}[h!]
	\centering
\hspace*{-1em}\includegraphics[scale=0.6]{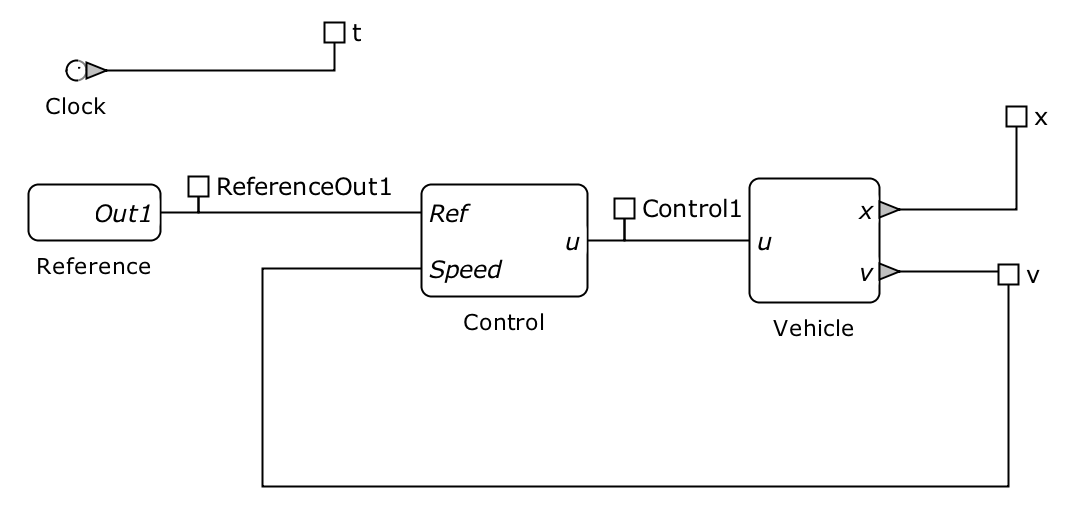}
	\caption{Cruise Control - Top Level - SpaceEx}
	\label{SXcar1}
\end{figure}

\begin{figure}[h!]
	\centering
\hspace*{-2em}\includegraphics[scale=0.6]{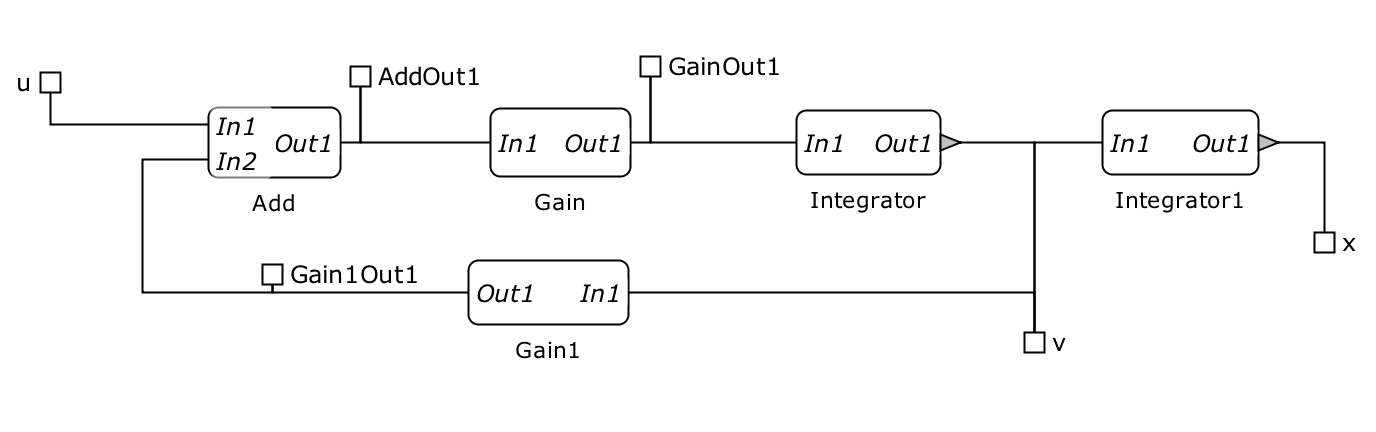}
	\caption{Cruise Control - Vehicle Dynamics - SpaceEx}
	\label{SXcar2}
\end{figure}

\begin{figure}[h!]
	\centering
\includegraphics[scale=0.53]{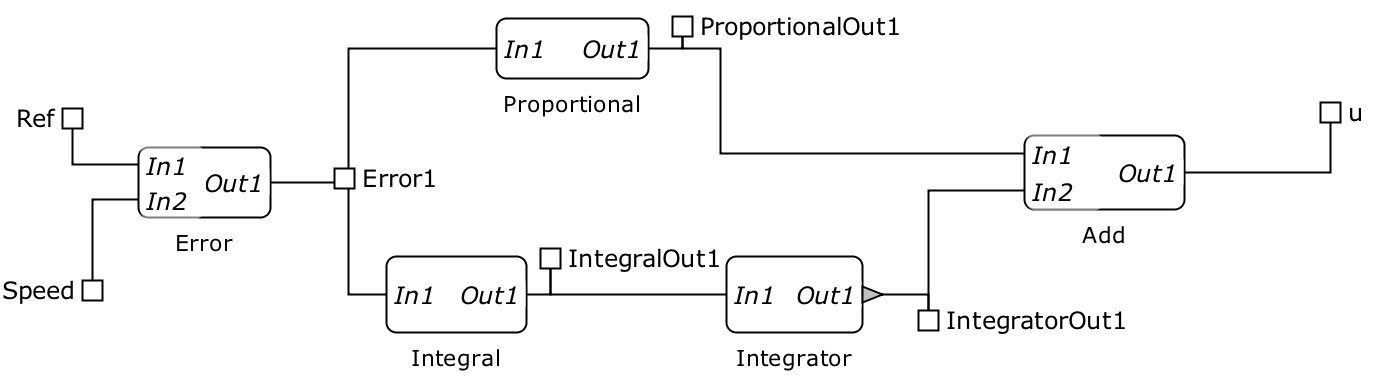}
	\caption{Cruise Control - Controller - SpaceEx}
	\label{SXcar3}
\end{figure}

In this Section, we create a SpaceEx (\texttt{CC\_SpaceEx.xml}) model for the cruise control system. We rely on the \texttt{SL2SX} translator~\cite{minopoli2016sl2sx} and since the blocks are linear we automatically get the complete SpaceEx model. For nonlinear or other blocks, we could use the techniques presented in~\cite{kekatos2017constructing,kekatos2018formal}.  The model consists of 12 base components (building blocks) and 2 network components (systems and subsystems). More information on SpaceEx (format, components, etc.) could be found at \cite{Intro}. The SpaceEx model is presented in Figures \ref{SXcar1}, \ref{SXcar2} and \ref{SXcar3}. %\unsure{I could add more information regarding the SpaceEx model, if you find it appropriate.}

\newpage

After constructing the SpaceEx model (XML), we need to create a configuration (CFG) file in which we select the initial conditions, parameters and scenarios. The \texttt{CC\_SpaceEx.cfg} file considers the STC scenario, an accuracy of 0.1, a global time horizon of 10s and an initial speed $0\leq v \leq2$. Now we are ready to run SpaceEx web platform. %\unsure{I could add more information about the SpaceEx platform and how we run a model with the web platform.} 

\begin{figure}[ht!]
	\centering
\includegraphics[scale=0.4]{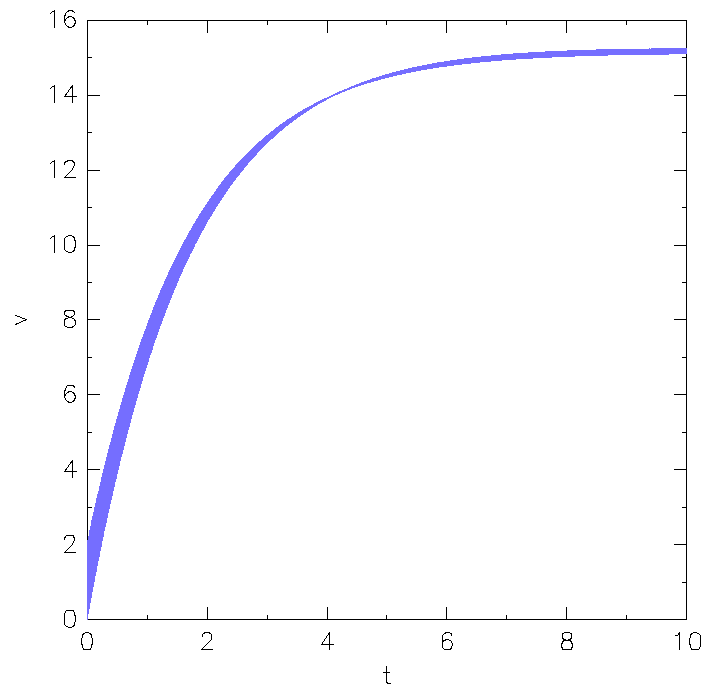}
	\caption{Reachable Sets with SpaceEx}
	\label{SX_CC}
\end{figure}

The reachable sets for the considered scenario are presented in Figure \ref{SX_CC}.

The requirements of steady-state error and rise-time are met. We can visually observe them too. The rise time (corresponding to 13.5) is less than 5 seconds. The speed stays in the range $[0.95*15, 1.05*15]= [14.25, 15.75]$. For more complicated safety properties, one can use the monitors described in~\cite{frehse2018toolchain}.
%\begin{thebibliography}{99}
%\bibitem{CPS}Radhakisan Baheti and Helen Gill, Cyber-physical Systems, \url{http://ieeecss.org/sites/ieeecss.org/files/documents/IoCT-Part3-02CyberphysicalSystems.pdf}.
%\bibitem{ref}Cruise Control Case Study,  \url{http://ctms.engin.umich.edu/CTMS/index.php?example=CruiseControl&section=ControlStateSpace}.
%\bibitem{Intro} Goran Frehse, An introduction to SpaceEx,
%\url{http://spaceex.imag.fr/sites/default/files/introduction_to_spaceex_0.pdf}.
%\bib
%\end{thebibliography}
	\bibliographystyle{abbrv}
	%\nocite{*}
	%\bibliography{refs}
	\bibliography{references}	
	\appendix
	\section*{Running SpaceEx -- Step-by-step guide}

	Briefly, the necessary steps to run SpaceEx are as follows.

\begin{itemize}
\item Run the virtual machine,
\item Select the SpaceEx Server,
\item Copy the indicated IP address (e.g. 192.168.56.101) in a browser\footnote{If an error occurs, please follow the guidelines of \url{http://spaceex.imag.fr/documentation/user-documentation/installing-spaceex-vm-server-17.}}, 
\item Navigate to Run SpaceEx tab (top-right),
\item On the model file (on the left), press "Choose file" and select the \texttt{CC\_SpaceEx.xml} file, there is 
\item On the configuration file (on the left), press "Load" and select the \texttt{CC\_SpaceEx.cfg} file,
\item Now, you are ready to press "Start" and visualize the reachable sets!
\end{itemize}
\end{document}